\documentclass[sigconf]{acmart}
\usepackage{booktabs}
\usepackage{array}
\usepackage{amsmath}
\usepackage{amsfonts}
\usepackage{multirow}
\AtBeginDocument{%
  }

\setcopyright{none}
\copyrightyear{2027}
\acmYear{2027}
\acmDOI{}
\acmConference[KDD '27]{The 33rd ACM SIGKDD Conference on Knowledge Discovery and Data Mining}{August 1--5, 2027}{San Jose, CA, USA}
\acmISBN{}

\begin{document}

\title{Hybrid GPU–CPU Retrieval for Personalized Search at Ultra-Large Scale}

\author{%
  Hao Fu \quad Jichao Sun \quad Baiting Zhu \quad Qiaoling Liu \quad Yan Shi \quad Cheng Lu \\
  Liu Liu \quad Yubo Wang \quad Xin Yao \quad Xiangyu Niu \quad Xu Dong \quad Wenhan Lyu \\
  Chiyao Shen \quad Yinjie Huang \quad Minglei Chen \quad Shuai Ding \quad Li Fan \quad Xiao Kong}
\affiliation{%
  \institution{Meta Platforms, Inc.}
  \city{Menlo Park}
  \state{California}
  \country{USA}}

\makeatletter
\gdef\authors{Hao Fu\and Jichao Sun\and Baiting Zhu\and Qiaoling Liu\and Yan Shi\and Cheng Lu\and Liu Liu\and Yubo Wang\and Xin Yao\and Xiangyu Niu\and Xu Dong\and Wenhan Lyu\and Chiyao Shen\and Yinjie Huang\and Minglei Chen\and Shuai Ding\and Li Fan\and Xiao Kong}
\makeatother

\renewcommand{\shortauthors}{Hao Fu et al.}

\begin{abstract}
Embedding-based retrieval on user-generated content at the trillion-document scale exposes a sharp conflict between two production demands: deep, expressive personalization for queries with rich user intent, and broad coverage of a massive inventory under fixed latency and resource budgets. We characterize this as the \emph{personalization--scale paradox}: hosting the full serving inventory in GPU memory is too resource intensive, while CPU compute cannot execute the same interaction-heavy model on the latency-critical path.

We present a hybrid GPU--CPU co-serving system that resolves the paradox through orchestration rather than a new model class. A high-depth GPU pathway fuses retrieval and interaction pre-ranking over a curated online pool on the order of a billion documents, while a high-breadth CPU pathway searches an independently selected online inventory roughly twenty times larger with lightweight personalized scoring. Either or both pathways can run per request; candidates are deduplicated before shared downstream ranking.

The system is deployed in production. A full-system A/B test against the legacy CPU-only configuration improves model-scored relevance and substantive engagement, while separate pathway experiments show positive value at their own deployment scopes. Retrieval logs show that the pathways contribute structurally distinct candidates, production serving measurements characterize their latency, and a matched capacity plan quantifies the economic rationale for assigning modeling depth to GPUs and inventory breadth to CPUs. Together, these results validate a practical, independently evolvable depth--breadth architecture for ultra-large-scale personalized search.
\end{abstract}





\maketitle

\section{Introduction}

Embedding retrieval over user-generated content depends on query semantics and evolving user interests. Discovery queries benefit from non-linear interactions with behavioral history, while navigational and specific queries require rare items from a broader inventory. At a multi-trillion-document \emph{source} corpus, this creates a \emph{personalization--scale paradox}: GPUs support deep interaction but cannot economically hold the full online inventory in high-bandwidth memory (HBM); CPUs provide affordable capacity but cannot run the same interaction model at scale.

User-generated content intensifies this tension: the same query can require different candidates as interests evolve, while new posts arrive and expired, deleted, or policy-ineligible items disappear. Serving must support changing user-conditioned scores and broad, frequently refreshed coverage; measuring only full-precision model quality or approximate nearest-neighbor (ANN) recall misses one side of the objective.

A discovery query may require behavioral history to prefer seafood over fast food, whereas an obscure restaurant-review query succeeds only if that sparse document remains retrievable. Modeling depth and inventory breadth consume different resources: interactions increase per-candidate compute, while rare content increases memory and scan cost. A homogeneous tier must compromise one side of this frontier.

We resolve this conflict with heterogeneous co-serving. A GPU pathway retrieves and pre-ranks over a curated inventory on the order of a billion documents, while a CPU pathway uses lightweight embedding matching over an inventory roughly twenty times larger, including popular content. Either or both can run per request and be disabled independently for capacity control or rollback. In a separate matched broad-vector plan, annualized accelerator capacity cost is roughly four times the CPU cost.

We study candidate generation and GPU interaction pre-ranking. Downstream full ranking, query-understanding training, generative retrieval, and learned request selection are out of scope; lexical retrieval appears only in measurements at the shared-ranker input. Our deployed architecture uses established two-tower and DeepFM-style components, validated through a full-system A/B test, pathway experiments, serving measurements, and source-attributed retrieval logs.

Concretely, this paper makes three contributions:
\begin{itemize}
    \item \textbf{Heterogeneous co-serving system.} We jointly serve an independently selected billion-scale GPU inventory and a CPU inventory roughly twenty times larger through separately versioned publication, deadlines, rollback, and a shared aggregation interface.
    \item \textbf{Depth--breadth pathway design.} The GPU pathway fuses two-tower retrieval with interaction scoring. The CPU pathway combines a dedicated embedding index, eager term-at-a-time scoring, bulk scanning, and finer clustering; component cost reductions are reported against their own baselines rather than multiplied together.
    \item \textbf{Production validation.} Separate pathway experiments, source-attributed retrieval logs, and the full-system A/B test establish pathway-level value, candidate diversity, and system-level value, respectively. We additionally report a conservative cross-day-dependence-robust GSRR interval, serving measurements, and a computational-cost comparison between GPU and CPU broad-inventory plans.
\end{itemize}

\section{Related Work}

\subsection{Personalized Retrieval and Interaction}
Deep retrieval spans DSSM~\cite{huang2013learning}, two-tower models~\cite{covington2016deep}, and graph-based industrial recommendation~\cite{ying2018graph,zhang2024lignn}. Language models have also improved representations and retrieval~\cite{wang2024improving,li2023llm}. These systems generally retain separable query--document scoring for efficient candidate generation. ColBERT~\cite{khattab2020colbert}, generative DSI~\cite{tay2022transformer}, and the target-aware generate--rank design of GRank~\cite{sun2025grank} increase expressiveness but make large-scale serving more demanding. We apply non-linear interaction scoring inside an accelerator-resident retrieval pass~\cite{silvertorch} before candidates leave the GPU.

\subsection{Industrial Scaling and Heterogeneous Systems}
Industrial examples include PinSage~\cite{ying2018graph}, Facebook search~\cite{huang2020embedding}, Alibaba commodity retrieval~\cite{wang2018billion}, and Que2Search~\cite{liu2021que2search}. They scale through hard-negative mining, real-time intent modeling~\cite{grbovic2018real}, and multi-tower architectures. SVFusion~\cite{tian2026svfusion} and FusionANNS~\cite{tian2024fusionanns} divide one ANN call across CPUs and GPUs. Our system instead operates distinct model families and independently selected inventories that meet at aggregation.

\subsection{Efficient Indexing and Tiered Serving}
IVF-PQ~\cite{jegou2010product}, HNSW~\cite{malkov2018efficient}, and DiskANN~\cite{subramanya2019diskann} optimize the cost--recall frontier of a single ANN engine. Disk-resident SPANN~\cite{chen2021spann} and SPFresh~\cite{xu2023spfresh} extend that frontier through tiered posting lists and online index maintenance, while GPU libraries such as Faiss~\cite{johnson2019billion} and cuVS~\cite{faiss_cuvs_2025} provide high scan throughput but remain bounded by HBM capacity. These advances are complementary to our contribution: either pathway could adopt a new index or kernel without changing the co-serving and candidate-aggregation contracts.

\begin{table}[t]
\centering
\scriptsize
\setlength{\tabcolsep}{3pt}
\caption{Architectural comparison at each paper's reported evaluation scale. Fused pre-rank denotes learned interaction scoring within the retrieval pass, before shared aggregation.}
\label{tab:related_compare}
\begin{tabular}{@{}p{0.27\columnwidth}ccc@{}}
\toprule
System & Scale & Hardware & Fused pre-rank \\
\midrule
Facebook search~\cite{huang2020embedding} & $100$M-scale index & CPU & No \\
DiskANN~\cite{subramanya2019diskann} & $1$B points & CPU & No \\
SPANN~\cite{chen2021spann} & $1$B vectors & CPU & No \\
SVFusion~\cite{tian2026svfusion} & Up to $1$B vectors & CPU+GPU & No \\
FusionANNS~\cite{tian2024fusionanns} & $1$B vectors & CPU+GPU & No \\
\midrule
\textbf{This work} & \textbf{Billion + tens-of-billions scale} & \textbf{CPU+GPU} & \textbf{Yes} \\
\bottomrule
\end{tabular}
\end{table}

Unlike SVFusion and FusionANNS, which co-process one ANN request, our independently selected billion- and tens-of-billions-scale inventories meet at aggregation. The GPU also pre-ranks candidates before aggregation. Section~\ref{sec:tco} compares broad-inventory capacity plans.

\section{System Architecture}
\label{sec:arch}

\begin{figure*}[t]
  \centering
  \includegraphics[width=0.95\textwidth,keepaspectratio]{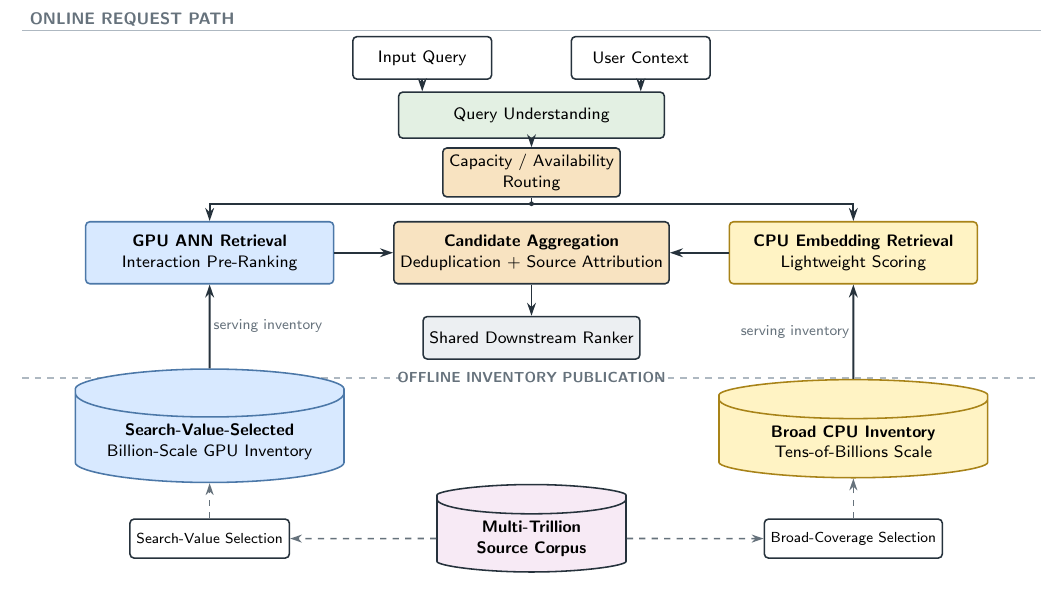}
  \caption{GPU--CPU co-serving over independently selected billion- and tens-of-billions-scale inventories from a multi-trillion-document source corpus.}
  \Description{An input query and user context feed a billion-scale GPU retrieval branch and a tens-of-billions-scale CPU retrieval branch selected from a multi-trillion-document source corpus. Their candidates are aggregated before downstream ranking.}
  \label{fig:arch}
\end{figure*}

Figure~\ref{fig:arch} separates the upstream source corpus from online retrieval: the multi-trillion figure describes the corpus before selection, not an ANN scan. After policy, quality, language, freshness, and duplication checks, hardware-specific selectors produce a GPU snapshot on the order of a billion documents and a CPU index roughly an order of magnitude larger. The CPU index is not a disjoint collection of low-popularity content; it also contains many popular documents eligible for the GPU pool. Because selection and publication are versioned independently, containment can vary.

\subsection{Parallel Retrieval and Aggregation}
Query understanding derives semantic, language, intent, and personalization signals from the query and available user context. Capacity and availability controls determine which pathways run; both can run for the same request. These controls support independent rollback and are not evaluated as a learned decision policy.

Active branches execute in parallel. The GPU branch combines ANN retrieval with interaction pre-ranking, while the CPU branch searches its dedicated embedding index. Each branch has its own deadline and can return independently. The aggregator deduplicates by document identifier and preserves source attribution before the shared downstream ranker.

The branches execute under independent deadlines rather than extending a serial critical path. Section~\ref{sec:rq3_overall} reports production latency for lexical retrieval and the CPU and GPU pathways; Section~\ref{sec:gpu_serving} separately reports the narrower per-accelerator model-server benchmark.

\subsection{Failure Isolation and Versioned Release}
The fork--join boundary also isolates failures. Each branch has a deadline; aggregation proceeds with candidates returned in time, so one slow branch does not discard the other's result. Releases are independent: the GPU publishes a matched model--index snapshot, while the CPU combines versioned centroids and a base index with live updates. Either pathway can be disabled without changing the aggregation interface.

\section{High-Depth GPU Retrieval}

The GPU pathway jointly trains retrieval and interaction scoring to prioritize modeling depth. This section follows its lifecycle from pool construction and multi-objective training to accelerator-resident retrieval and pre-ranking.

\begin{figure}[t]
  \centering
  \includegraphics[width=0.92\columnwidth,keepaspectratio]{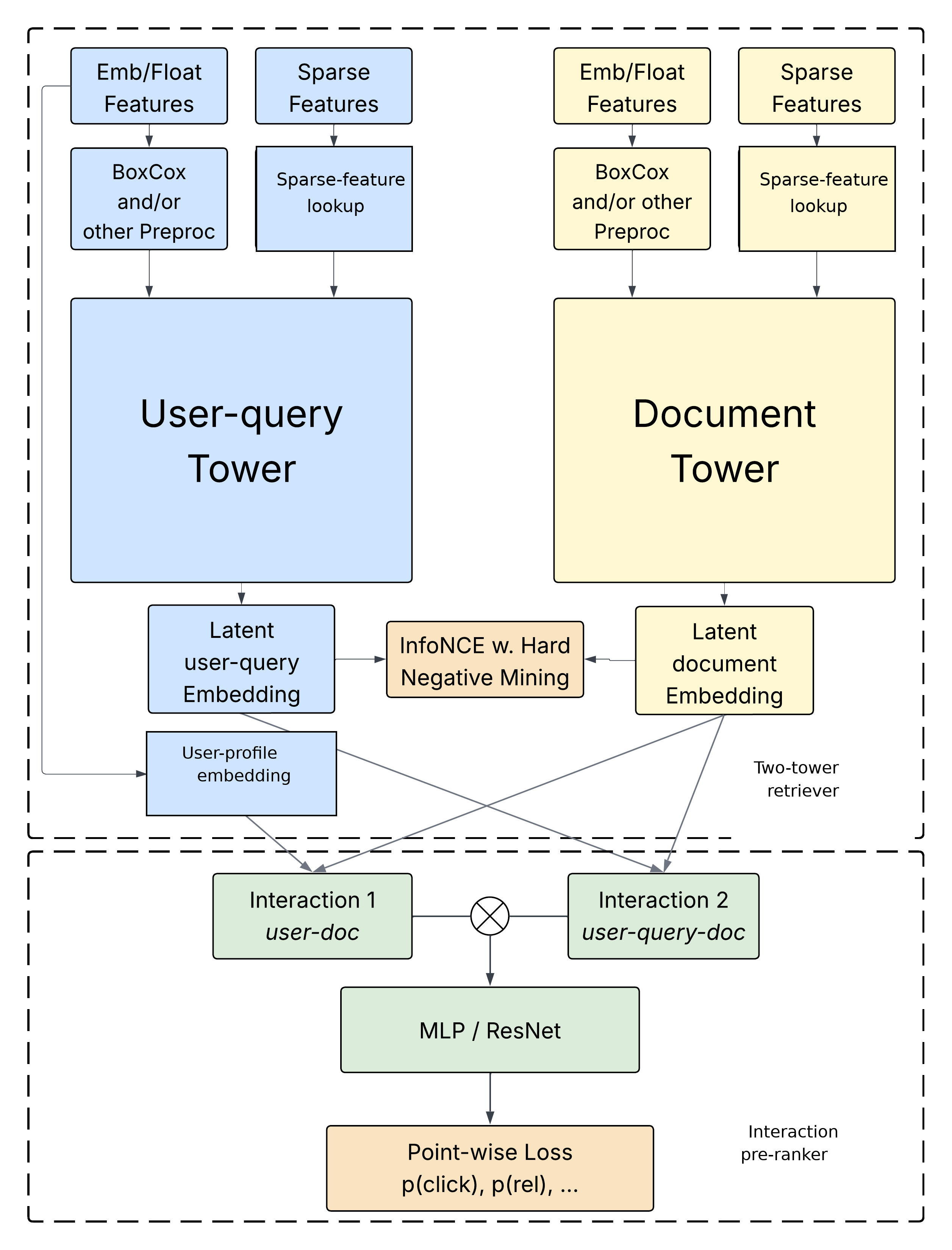}
  \caption{Joint training combines InfoNCE retrieval with engagement and relevance objectives.}
  \Description{User-query and document towers are jointly trained with an interaction module using retrieval, relevance, and engagement losses.}
  \label{fig:gpu_train}
\end{figure}

\subsection{Construction of the High-Quality Pool}
\label{sec:pool_construction}

Because high-bandwidth memory cannot hold the full inventory, the GPU pool optimizes \emph{search value}, not raw virality. Recommendation engagement is often dominated by entertainment with little query specificity, while local guides and tutorials may be essential to explicit search intents despite modest aggregate engagement. The selector protects this informative content while pruning viral items with low search utility.

\paragraph{Lifecycle-aware filtration.} Content passes through three stages as it ages. At ingest, hard metadata gates enforce language, originality, and content type. During the first week, early-engagement filters remove spam and low-value items. From day~7 onward, taxonomy, search-usefulness, local-information, and temporal-decay models estimate long-term utility. Staging prevents mature-content thresholds before new items accrue evidence.

\paragraph{Adaptive retention.} Decay is category-dependent: niche tutorials and local guides with high search value bypass engagement pruning, while news and trends decay aggressively. A highly selective evergreen pool retains under $0.1\%$ of content older than one year. The policy keeps the billion-scale pool fresh without discarding low-frequency content that remains useful for explicit search intents.

\subsection{Model Architecture and Joint Training}
The two-tower retriever and interaction pre-ranker are jointly trained on search logs, using InfoNCE and hard negatives to align semantic relevance and engagement.

Figure~\ref{fig:gpu_train} combines user-query and document towers with a DeepFM-based interaction module~\cite{factorization}. The query tower forms a 512-dimensional dense input from a 384-dimensional semantic embedding and a 128-dimensional user-profile embedding. A residual multilayer perceptron processes this input with sparse categorical context. The document tower consumes dense, categorical, and content-embedding features and emits a 128-dimensional vector for nearest-neighbor retrieval. Joint training aligns retrieval and scoring distributions.

Training uses three weighted objectives: contrastive InfoNCE for retrieval, Smooth L1 for relevance, and binary cross-entropy (BCE) for engagement,
\begin{equation}
    \mathcal{L}_{\text{Total}} = w_1 \mathcal{L}_{\text{InfoNCE}} + w_2 \mathcal{L}_{\text{SmoothL1}} + w_3 \mathcal{L}_{\text{BCE}},
\end{equation}
with hyperparameters $w_1, w_2, w_3$. Let $h(q,d)=\exp(\operatorname{sim}(q,d)/\tau)$. For a batch of size $B$ with anchor query $q_i$, positive document $d_i^+$, and negative documents $D_j$ drawn from every other session $j \neq i$, the two-tower stack uses the in-batch loss
\begin{equation}
\label{eq:infonce}
\mathcal{L}_{\text{InfoNCE}} = -\frac{1}{B} \sum_{i=1}^{B} \log
\frac{h(q_i,d_i^+)}{h(q_i,d_i^+)+\sum_{j\neq i}\sum_{d\in D_j}h(q_i,d)}.
\end{equation}
The interaction pre-ranker predicts relevance and engagement from impression labels or later-stage ranker distillation, using binary cross-entropy for engagement and Smooth L1 for relevance. In addition to the cross-session loss in Eq.~\eqref{eq:infonce}, a within-session InfoNCE term treats engaged candidates as positives and relevant but non-engaged candidates as negatives. These harder session-local negatives already match the query.

A value model computes $S_{\text{final}}=\sum_t\lambda_t\hat{p}_t$. The per-task weights balance relevance and engagement objectives and can vary by traffic segment. Consequently, one trained model can serve heterogeneous traffic and test new operating points without retraining.
Appendix~\ref{app:objectives} summarizes objective supervision and serving roles.

\paragraph{Training lifecycle.} The pre-ranker trains continuously from daily refreshed search logs. An automated pipeline publishes model weights and the candidate index together, so both reflect the same corpus view. Scheduled retraining is independent of capacity configuration, allowing an experiment to change the active operating point without rebuilding the model.

\paragraph{Index freshness.} The daily cadence matches the application's freshness requirements rather than a hardware limit. It also aligns the model checkpoint with the training-data view of the corpus. Other applications can choose a different cadence without changing the model or aggregation interface.

\subsection{Accelerator-Resident Serving}
\label{sec:gpu_serving}
\begin{figure}[t]
  \centering
  \includegraphics[width=0.92\columnwidth,keepaspectratio]{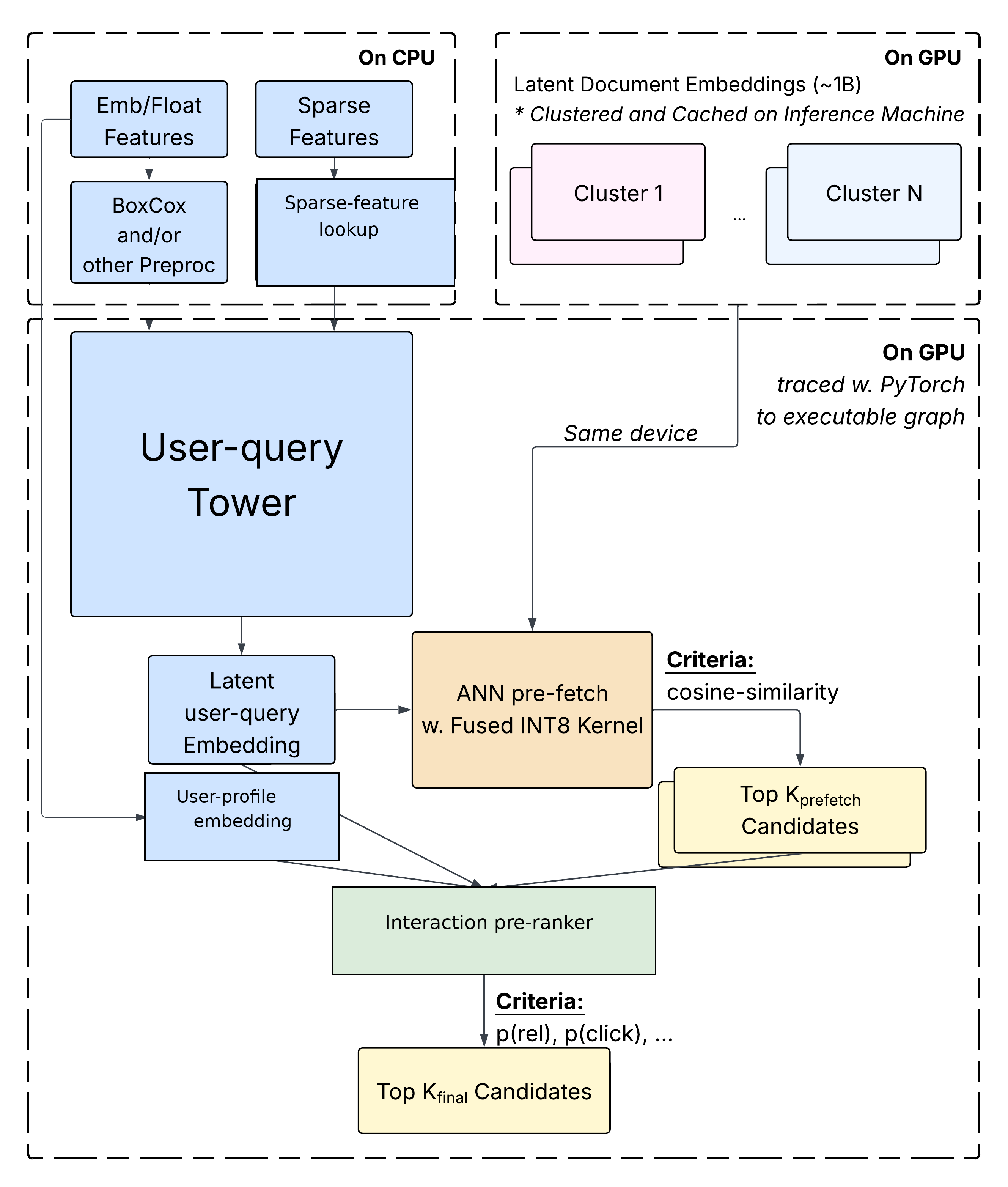}
  \caption{GPU inference; sparse tables remain on CPU.}
  \Description{A GPU-resident pipeline performs query encoding, approximate nearest-neighbor retrieval, interaction pre-ranking, and value scoring while fetching sparse tables from host memory.}
  \label{fig:gpu_inference}
\end{figure}

The serving stack follows SilverTorch~\cite{silvertorch} and keeps ANN search, inference, and pre-ranking accelerator-resident, avoiding the CPU--GPU transfer that would otherwise separate retrieval from scoring. Figure~\ref{fig:gpu_inference} shows the interaction pre-ranker filtering retrieved candidates before the value model combines its predictions. Candidate vectors and interaction features remain on the accelerator through the compute-intensive stages.

The compute backbone resides in high-bandwidth memory, while sparse embedding tables are fetched on demand from host memory over fifth-generation PCI Express (PCIe Gen5). This hybrid memory hierarchy lets model parameters exceed accelerator-memory capacity without moving latency-critical dense kernels off the GPU. Fused 8-bit integer kernels combine quantization and distance computation in one operator, reducing intermediate materialization and memory bandwidth during coarse candidate selection.

The deployment uses AMD MI300X accelerators. In a production load test, one card sustained roughly 150--200 queries per second at approximately 30--40~ms 99th-percentile model-server latency. These per-accelerator figures exclude network, aggregation, and downstream ranking.

\section{Broad-Inventory CPU Retrieval}

The CPU pathway complements GPU modeling depth with an order-of-magnitude larger inventory and a faster update loop. It trades away deep per-document interaction but preserves personalized semantic retrieval across a broad inventory that cannot be hosted economically in HBM.

\subsection{CPU Retrieval Architecture}
\label{sec:cpu_arch}

Pretrained centroids key inverted embedding lists. A query first selects nearby centroids for coarse pruning, then scans the selected lists and combines cosine similarity with lightweight lexical signals. A dedicated document-understanding service consumes both the real-time create/update stream and the static pool, dispatches content for embedding generation, and loads the resulting vectors into the serving index. Embedding generation is therefore decoupled from CPU retrieval even though the index itself resides in DRAM.

\subsection{Dedicated Embedding Index}
\label{sec:cpu_index}

Embedding search was previously confined to subsets of lexical indices. Sharing that substrate forced vector scoring through term-oriented abstractions and made its capacity contend with latency-sensitive lexical workloads. A dedicated embedding index reduces this coupling and expands the pool by more than an order of magnitude.

\paragraph{Document selection.}
The dedicated index holds tens of billions of public documents selected from the multi-trillion-document upstream corpus. Selection covers the principal languages in the target market, recent and historically useful content, and minimum quality requirements. Documents, embeddings, and a minimal lexical signal set are partitioned and replicated in dynamic random-access memory (DRAM) on commodity x86 servers. A batch builds the base index, while an online service applies updates.

\paragraph{Distributed cluster training.}
The $10\times$ larger pool requires proportionally more centroids to control list size. The previous indexing-time procedure trained 64k centroids separately per shard; extending the same serial workflow to 512k takes prohibitively long and blocks deployment. We decouple training from indexing: an independent distributed trainer runs daily across many machines, materializes a versioned centroid index, and publishes it for serving shards to ingest. Index servers no longer repeat clustering. Section~\ref{sec:cpu_eval} evaluates the 64k$\to$512k operating point.

\subsection{Efficient Serving: Eager Evaluation and Hit Scanning}
\label{sec:cpu_serving}

Legacy document-at-a-time (DAAT) execution thrashes cache on the tens-of-billions-scale pool because vector scoring jumps among document identifiers scattered across posting lists. We use two complementary optimizations.

\begin{figure}[t]
  \centering
  \includegraphics[width=0.96\columnwidth,keepaspectratio]{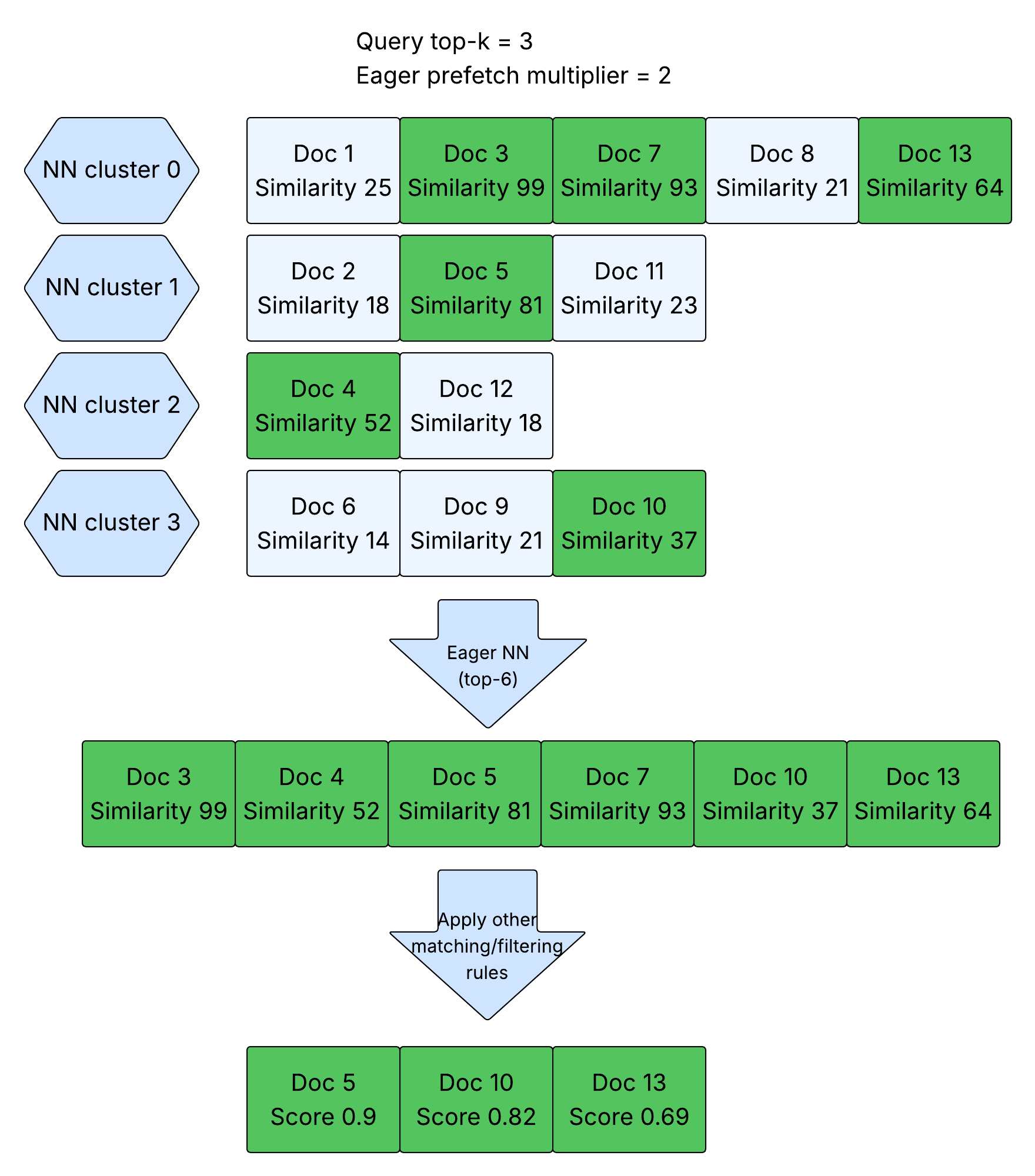}
  \caption{Eager retrieval for three nearest neighbors with prefetch multiplier $2$ scans clusters before applying other rules to the top-$6$.}
  \Description{Three nearest-neighbor clusters are scanned sequentially to produce six prefetched candidates before other query constraints are evaluated.}
  \label{F:eager}
\end{figure}

\paragraph{Eager evaluation.}
We invert the query plan instead of interleaving nearest-neighbor scoring with Boolean constraints. The system first scans relevant nearest-neighbor clusters term-at-a-time (TAAT) with contiguous reads that maximize cache locality, computes a global Top-$K$, and then uses that candidate list as a dynamic skip list for the remaining DAAT constraints. Heavy vector computation therefore completes before Boolean processing, converting scattered random reads into sequential scans and preventing computationally expensive distance calculations on documents that cannot enter the prefetched set (Figure~\ref{F:eager}).

\paragraph{Hit scanning.}
The iterator-based hit abstraction incurs a virtual-function call for every matched document. We bypass it by passing raw vectors in bulk to 512-bit vector-instruction (AVX-512) list-scanning kernels, shifting cycles from dispatch and pointer chasing to useful distance computation.

Both optimizations run on the dedicated embedding index, so high-throughput vector scans do not contend with latency-sensitive lexical operations or inherit their iterator overhead.

\subsection{Lightweight Personalized Modeling}
\label{sec:cpu_modeling}

The CPU model applies lightweight contextual personalization and multilingual semantics to the tens-of-billions-scale inventory. Its purpose is not to reproduce GPU interaction depth, but to retain a useful user-conditioned representation under a much lower per-candidate compute budget.

\paragraph{Unified two-tower model.} User-query and document towers share a fine-tuned XLM-V backbone~\cite{liang2023xlmv}, trained with bidirectional InfoNCE to produce compact 384-dimensional multilingual vectors. An attention module fuses the dense query representation with sparse contextual features through a pooled embedding lookup. The shared backbone reuses document embeddings across contexts while keeping personalization query-side.

\paragraph{Scale trade-offs.} User and document vectors interact only by dot product, excluding the cross-feature interactions used by the GPU pathway. Features are limited to lightweight profile, coarse context, and history signals. These restrictions keep document vectors precomputable and make scans suitable for single-instruction, multiple-data execution. The deliberate loss of expressiveness buys inventory breadth: the CPU pathway retrieves from tens of billions of documents, while later stages apply richer modeling only to the much smaller surviving set.

\section{Evaluation}

We ask four questions. \textbf{RQ1}: What is the end-to-end value of the high-depth GPU pathway? \textbf{RQ2}: Can the CPU pathway serve a tens-of-billions-scale inventory efficiently and improve online quality? \textbf{RQ3}: Does the combined system improve production quality while contributing distinct candidates? \textbf{RQ4}: For a fixed broad-vector workload, what capacity and computational cost follow from GPU versus CPU placement?

\subsection{Experimental Setup and Metrics}

We evaluate the system through production A/B tests, serving benchmarks, and retrieval logs. The headline comparison is a full-system A/B test against the legacy CPU-only configuration. It uses persistent account-level assignment, split approximately equally between control and treatment. Daily assigned populations are on the order of millions of accounts per arm. The seven-day GPU-pathway experiment and nine-day CPU-pathway evaluation with the GPU pathway disabled ran separately against contemporaneous production controls.

\paragraph{Metrics.}
\textbf{DCG@20} is model-scored relevance over the top 20 results, not a human-rater measure; we interpret it with behavioral outcomes rather than as independent ground truth.

\textbf{GSRR (Good Search Result Rate)} is a retention-oriented metric of substantive engagement. For the eligible search sessions $\mathcal{S}$, let $\mathcal{E}_s$ be the events observed in session $s$, $v(e)$ the strength of event $e$, and $\tau_{\operatorname{type}(e)}$ its event-type-specific threshold. We define
\begin{align}
g(s)&=\mathbb{1}\!\left[\exists e\in\mathcal{E}_s:
v(e)\geq\tau_{\operatorname{type}(e)}\right],\\
\mathrm{GSRR}&=\frac{1}{|\mathcal{S}|}\sum_{s\in\mathcal{S}}g(s).
\end{align}
Thus, a good session contains at least one qualifying event, such as a sufficiently long view or an explicit positive action.

\paragraph{Statistical reporting.}
All reported metric changes are relative lifts, computed as $100(\mu_t/\mu_c-1)$ from the treatment and control means. Statistical uncertainty is computed at the account-assignment level, so sessions from the same account are not treated as independent. For full-system GSRR, we give equal weight to each of the nine consecutive daily lifts. Persistent assignment also reuses accounts across days, so we do not treat daily estimates as independent or divide their uncertainty by $\sqrt{9}$. Instead, for each day we take the larger side of its reported 95\% interval and average the nine half-widths. By the Cauchy--Schwarz bound, the resulting half-width is conservative under arbitrary cross-day dependence, including perfect correlation. It yields the rounded 95\% interval $[+1.71\%,+2.32\%]$; all nine daily intervals are also above zero. The $\pm$ values for the separate pathway experiments below are their two-sided 95\% interval half-widths. All reported lifts and intervals are rounded to two decimal places.

\begin{table}[t]
\centering
\footnotesize
\setlength{\tabcolsep}{3pt}
\caption{Production evaluation. Relative lifts use contemporaneous controls; experiments ran in separate windows and should not be compared across rows. For the hybrid row, both intervals are the conservative nine-day interval described in Appendix~\ref{app:repro}.}
\label{tab:main_results}
\begin{tabular}{@{}p{0.30\columnwidth} >{\centering\arraybackslash}p{0.24\columnwidth} >{\centering\arraybackslash}p{0.34\columnwidth}@{}}
\toprule
Configuration & DCG@20 & GSRR \\
\midrule
High-depth GPU & $+0.73\%\pm0.15\%$ & $+1.04\%\pm0.13\%$ \\
Broad-inventory CPU & $+3.16\%\pm0.15\%$ & $+1.20\%\pm0.11\%$ \\
\midrule
\textbf{Hybrid architecture} & \textbf{$+4.51\%$ $[+3.99\%,+5.04\%]$} & \textbf{$+2.01\%$ $[+1.71\%,+2.32\%]$} \\
\bottomrule
\end{tabular}
\end{table}

Over the same nine-day window, the hybrid treatment improved the equal-day mean DCG@20 by $4.51\%$ and the equal-day mean GSRR by $2.01\%$ relative to the legacy all-CPU configuration. The separate pathway experiments establish positive online value for each pathway against its contemporaneous production control. Candidate-set analysis shows that they contribute structurally distinct results, while the full-system A/B test validates the deployed co-serving operating point. Together, these results demonstrate complementarity in the operational sense.

\paragraph{Evaluation scope.}
The production tests evaluate complete pathways under real-time state, policy filters, and a shared downstream ranker rather than isolated ANN algorithms. Because each treatment changes multiple components, its online lift measures the complete treatment configuration; attribution to a kernel, inventory change, or model layer requires a corresponding isolated experiment.

\subsection{GPU Pathway: End-to-End Modeling Value (RQ1)}
\label{sec:gpu_eval}

A predecessor semantic-model deployment illustrates why offline model quality alone is insufficient. It improved offline relevance but regressed online viewing. Investigation identified three deployment mismatches: offline evaluation used exact nearest-neighbor search while serving used a quantized ANN configuration; offline document text was richer than the online representation; and a downstream filter remained calibrated to the previous model. The embedding, ANN configuration, online features, and downstream filters therefore form one deployment unit.

The deployed GPU pathway changes accelerator serving, the ANN operating point, and interaction scoring together, so Table~\ref{tab:main_results} reports an end-to-end comparison. The seven-day experiment measured $+1.04\%\pm0.13\%$ GSRR and $+0.73\%\pm0.15\%$ DCG@20. A production load test measured on the order of ten billion floating-point operations per query (roughly twice that at peak) at a per-card throughput of a few hundred queries per second. This compute profile makes interaction pre-ranking practical but does not isolate hardware, model, or ANN effects.

\subsection{CPU Pathway: Scale and Efficiency (RQ2)}
\label{sec:cpu_eval}

With the GPU pathway disabled, the nine-day CPU-only evaluation measured the lightweight CPU personalization model together with a relevance-filter update. It delivered $+1.20\%\pm0.11\%$ GSRR and $+3.16\%\pm0.15\%$ DCG@20 (Table~\ref{tab:main_results}). This is a complete CPU-pathway estimate: it does not isolate inventory breadth, the personalization model, the dedicated index, or the filter update.

\begin{figure}[t]
  \centering
  \includegraphics[width=\columnwidth,keepaspectratio]{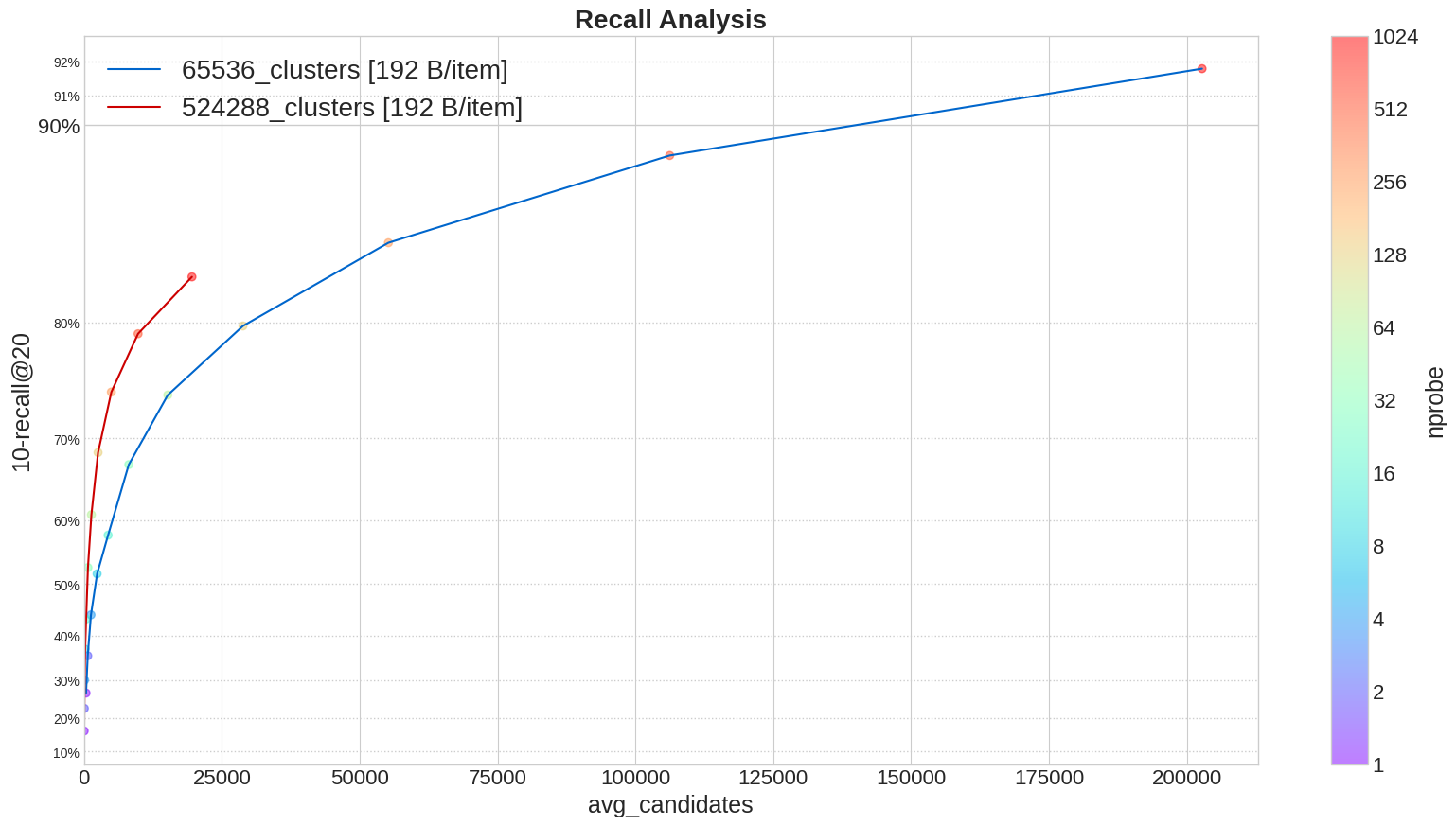}
  \caption{Recall--candidate frontier for 64k and 512k centroid indices as the number of probed clusters varies from 1 to 1024. The 512k index reaches comparable recall with fewer scanned candidates.}
  \Description{Recall at 20 versus average scanned candidates for 64k- and 512k-centroid indices. Point color encodes the number of probed clusters. The 512k curve reaches the same recall with fewer candidates over their overlapping range.}
  \label{fig:centroid_recall}
\end{figure}

Dedicated indexing and eager execution reduced retrieval-stage compute by 89.45\% (9.48$\times$). Figure~\ref{fig:centroid_recall} separately shows the offline recall--candidate frontier for 64k and 512k centroid indices. A 15-day online experiment then evaluated the selected 512k-centroid configuration with 256 probed clusters ($n_{\mathrm{probe}}=256$): it reduced CPU use by 63.5\% without a detected GSRR regression. We report the two reductions separately because their baselines differ and each measurement includes fixed overhead.

\subsection{Overall Production Value and Candidate Diversity (RQ3)}
\label{sec:rq3_overall}

The production A/B directly compares the deployed hybrid system with its legacy all-CPU predecessor. Over nine consecutive days the hybrid treatment improves the equal-day mean of model-scored DCG@20 by $4.51\%$ (conservative 95\% confidence interval $[+3.99\%,+5.04\%]$) and of GSRR by $2.01\%$ ($[+1.71\%,+2.32\%]$; Table~\ref{tab:main_results}). Every daily interval of both metrics is above zero. These headline estimates validate the complete production operating point; they do not identify an interaction effect or assign the lift among hardware, inventory, models, eligibility controls, and aggregation.

\paragraph{Production retrieval-branch latency.}
Table~\ref{tab:retrieval_latency} reports seven days of production wall time from entry to exit of each post-retrieval branch at the top-level aggregator. We restrict log records to post retrieval and apply the logging system's sampling weights. For the CPU pathway, we combine all dense retrieval work belonging to the same retrieval call before computing percentiles; the displayed values are therefore not averages of per-source percentiles. Retrieval for short-form video and other video surfaces is excluded.

\begin{table}[t]
\centering
\footnotesize
\setlength{\tabcolsep}{2pt}
\caption{Production retrieval-branch wall time for posts over seven complete days. Call counts and latency percentiles are sampling-weighted and measured at the retrieval-branch boundary.}
\label{tab:retrieval_latency}
\begin{tabular}{@{}lrrrr@{}}
\toprule
Branch & Calls & P50 (ms) & P95 (ms) & P99 (ms) \\
\midrule
Lexical & 3.43M & 196 & 686 & 1{,}184 \\
CPU pathway & 2.11M & 41 & 439 & 859 \\
GPU pathway & 0.94M & 151 & 675 & 1{,}376 \\
\bottomrule
\end{tabular}
\end{table}

The CPU pathway is faster than lexical retrieval at all three reported percentiles. The GPU pathway is faster at P50 and P95, while its P99 is 16\% higher. This is not a like-for-like kernel comparison: the GPU pathway applies interaction pre-ranking and passes approximately 224 candidates to the shared-ranker input, versus 59 from the dense CPU pathway (Table~\ref{tab:dense_flow}). The distributions therefore characterize two deployed operating points with different candidate budgets and scoring depth, rather than ranking one pathway's computational efficiency. They describe current production branch execution, not the historical A/B window or complete-search latency, and exclude query understanding, network transit outside the measured branch boundary, candidate aggregation, downstream ranking, and client rendering. The independent benchmark in Section~\ref{sec:gpu_serving} reports a tens-of-milliseconds P99 for the accelerator model server alone and is narrower still.

\paragraph{Candidate-set analysis.}
\label{sec:pathway_overlap}
For each search request in a three-day log window for which both dense pathways ran, let $G$ be the GPU retrieval set and $C$ the union of three dense CPU retrieval sets. We compute containment and Jaccard per request and average within query-popularity segments. Head is the top popularity decile, torso is deciles 2--5, and tail is deciles 6--10. Table~\ref{tab:overlap} shows low overlap across approximately 114k eligible requests.

\begin{table}[t]
\centering
\small
\setlength{\tabcolsep}{3pt}
\caption{Macro-averaged GPU--CPU dense-retrieval overlap over three days. GPU-side is $|G\cap C|/|G|$; CPU-side is $|G\cap C|/|C|$.}
\label{tab:overlap}
\begin{tabular}{@{}lrrr@{}}
\toprule
Segment (queries) & GPU-side & CPU-side & Jaccard \\
\midrule
Head ($85{,}617$)  & $2.41\%$ & $1.15\%$ & $0.75\%$ \\
Torso ($17{,}057$) & $2.08\%$ & $0.60\%$ & $0.44\%$ \\
Tail ($11{,}405$)  & $2.20\%$ & $0.52\%$ & $0.41\%$ \\
\bottomrule
\end{tabular}
\end{table}

Low overlap establishes candidate diversity, not user value by itself. The production system A/B supplies the overall user-value evidence, while the separate pathway launches show gains for the complete deployed pathways under their own controls. Composition at the shared-ranker input provides a second structural check: the pool averages approximately 781 deduplicated candidates per eligible request, with independently rounded group averages of 209 GPU-only, 47 dense-CPU-only, 500 lexical-CPU-only, and 26 cross-group candidates. These counts do not establish final Top-20 exposure or per-source engagement attribution; Appendix~\ref{app:overlap} gives definitions and robustness checks.

\paragraph{Contribution at the shared-ranker input.}
Both dense pathways continue to contribute after their own scoring stages. On the same three-day slice, the GPU pathway retrieves an average of 594 candidates and passes 224 to the shared-ranker input. The deduplicated dense CPU union retrieves 1,352 and passes 59 (Table~\ref{tab:dense_flow}). Because their candidate budgets and filtering stages differ, the reach rates describe two candidate pipelines rather than rank pathway quality. GPU interaction scoring selects from the curated billion-scale pool under a precision-oriented budget, while CPU retrieval starts with a roughly twenty-times-larger inventory and applies a lighter filter. The production A/B tests supply the user-value evidence.

\begin{table}[t]
\centering
\small
\setlength{\tabcolsep}{4pt}
\caption{Dense-pathway candidate flow over three days. Counts are rounded averages over eligible content-search requests for which the GPU pathway ran; reach rates use unrounded counts. The CPU row is the deduplicated union of all dense CPU sources.}
\label{tab:dense_flow}
\begin{tabular}{@{}lrrr@{}}
\toprule
Pathway & Retrieved & Ranker input & Reach rate \\
\midrule
GPU pathway & 594 & 224 & $37.6\%$ \\
Dense CPU union & 1,352 & 59 & $4.4\%$ \\
\bottomrule
\end{tabular}
\end{table}

The result is also stable to a longer window. A separate seven-day check against the same dense CPU union gives GPU-side overlap of $2.32\%$, $2.07\%$, and $2.15\%$ for head, torso, and tail, within 0.13 percentage points of Table~\ref{tab:overlap}.

\subsection{Broad-Inventory Capacity Cost (RQ4)}
\label{sec:tco}

The deployed CPU inventory is at tens-of-billions scale; this capacity plan is separate. It compares GPU and CPU placement for the same broad vector-search workload using 256-dimensional vectors encoded with scalar 8-bit quantization (SQ8). Each alternative uses three full regional replicas for disaster recovery and satisfies the same regional throughput requirement. Table~\ref{tab:tco_main} reports normalized planning inputs rather than absolute fleet counts.

\begin{table}[t]
\centering
\scriptsize
\setlength{\tabcolsep}{4pt}
\caption{Normalized capacity-plan inputs for a matched $d=256$ SQ8 workload with three regional replicas. Capacity-cost values use the CPU plan as the reference.}
\label{tab:tco_main}
\begin{tabular}{@{}lrr@{}}
\toprule
Planning input & CPU host & Accelerator \\
\midrule
Unit capacity cost & $1.0\times$ & $\sim12\times$ \\
Query throughput/unit & $1.0\times$ & $\sim3$--$4\times$ \\
Vector capacity/unit & $1.0\times$ & $\sim2$--$3\times$ \\
Units/region & $1.0\times$ & $\sim1/3\times$ \\
3-region capacity cost & $1.0\times$ & $\sim4\times$ \\
\bottomrule
\end{tabular}
\end{table}

Storage, rather than the throughput floor, determines both unit counts. The accelerator stores roughly two to three times as many vectors per unit and therefore uses roughly one-third as many units per region. After normalizing the CPU plan to $1.0\times$, the rounded planning inputs give an accelerator unit-capacity-cost ratio of approximately $12\times$ and a three-region accelerator/CPU capacity-cost ratio of roughly $4\times$. The comparison excludes shared ranking, networking, the deployed billion-scale GPU pathway, and utilization-dependent overhead. It also does not establish latency or retrieval-quality equivalence between the two implementations; it is an auditable capacity-plan result for the stated vector representation, not total capacity cost for the deployed hybrid system. Appendix~\ref{app:tco} details the plan's scope and assumptions.

\section{Discussion and Future Work}
\label{sec:discussion}

The stable interface in this deployment is the candidate contract: each pathway owns its inventory, model, eligibility, publication cadence, deadline, and rollback, while the aggregator owns source attribution and deduplication. The failed predecessor in Section~\ref{sec:gpu_eval} shows why an embedding alone is not a release unit: the online representation, ANN operating point, filters, model, and index must be versioned as a compatible bundle. Evidence also needs matched scope. The full-system A/B establishes overall value, pathway experiments estimate their complete treatment configurations, and overlap logs test structural redundancy without substituting for causal evidence. Likewise, the capacity comparison fixes representation, inventory, throughput, and replication; it informs one placement decision rather than whole-system capacity cost.

The pathway experiments change multiple components, so their gains cannot be assigned to hardware, ANN precision, model depth, or inventory size individually. Learned request-level routing is a direct next step, but unbiased evaluation requires randomized pathway assignment with quality, deadline, and capacity-cost logging. Finally, shared-ranker input diversity does not reveal source-attributed Top-20 exposure or engagement; those outcomes require attributed impressions and actions.

\section{Conclusion}

We presented a deployed depth--breadth retrieval architecture that treats model expressiveness and inventory coverage as separate serving objectives. A GPU pathway performs interaction-heavy retrieval over a curated inventory, while a CPU pathway supplies broad semantic coverage. Independent selection, publication, deadlines, and rollback allow both pathways to co-serve behind a stable candidate contract and evolve without forcing one implementation to follow the other. Production experiments validate the combined operating point, pathway launches establish value at their own deployment scope, and retrieval logs show that the pathways contribute structurally distinct candidates. The broader lesson is that ultra-large-scale personalized retrieval need not force one hardware tier to optimize incompatible objectives: heterogeneous pathways can specialize, evolve independently, and be evaluated as production bundles while preserving explicit quality, latency, and capacity-cost trade-offs.

\section*{Ethical Considerations}
Deployment and controlled experiments followed institutional privacy and experimentation-review processes. They added no data collection or direct recruitment; models use existing access-controlled signals, and we report aggregates under retention controls. These aggregates do not establish subgroup parity; subgroup quality and creator exposure remain limitations.

\section*{GenAI Usage Disclosure}
GenAI assisted source discovery and editing; the authors verified all claims and remain responsible for the manuscript.

\clearpage
\bibliographystyle{ACM-Reference-Format}
\bibliography{ref}


\begin{thebibliography}{25}


\ifx \showCODEN    \undefined \def \showCODEN     #1{\unskip}     \fi
\ifx \showISBNx    \undefined \def \showISBNx     #1{\unskip}     \fi
\ifx \showISBNxiii \undefined \def \showISBNxiii  #1{\unskip}     \fi
\ifx \showISSN     \undefined \def \showISSN      #1{\unskip}     \fi
\ifx \showLCCN     \undefined \def \showLCCN      #1{\unskip}     \fi
\ifx \shownote     \undefined \def \shownote      #1{#1}          \fi
\ifx \showarticletitle \undefined \def \showarticletitle #1{#1}   \fi
\ifx \showURL      \undefined \def \showURL       {\relax}        \fi
\providecommand\bibfield[2]{#2}
\providecommand\bibinfo[2]{#2}
\providecommand\natexlab[1]{#1}
\providecommand\showeprint[2][]{arXiv:#2}

\bibitem[Borisyuk et~al\mbox{.}(2024)]%
        {zhang2024lignn}
\bibfield{author}{\bibinfo{person}{Fedor Borisyuk} {et~al\mbox{.}}} \bibinfo{year}{2024}\natexlab{}.
\newblock \showarticletitle{LiGNN: Graph Neural Networks at LinkedIn}.
\newblock \bibinfo{journal}{\emph{arXiv preprint arXiv:2402.11139}}  \bibinfo{volume}{abs/2402.11139} (\bibinfo{year}{2024}), \bibinfo{pages}{1--12}.
\newblock


\bibitem[Chen et~al\mbox{.}(2021)]%
        {chen2021spann}
\bibfield{author}{\bibinfo{person}{Qi Chen}, \bibinfo{person}{Bing Zhao}, \bibinfo{person}{Haidong Wang}, \bibinfo{person}{Mingqin Li}, \bibinfo{person}{Chuanjie Liu}, \bibinfo{person}{Zengzhong Li}, \bibinfo{person}{Mao Yang}, {and} \bibinfo{person}{Jingdong Wang}.} \bibinfo{year}{2021}\natexlab{}.
\newblock \showarticletitle{SPANN: Highly-efficient billion-scale approximate nearest neighborhood search}. In \bibinfo{booktitle}{\emph{Advances in Neural Information Processing Systems (NeurIPS)}}. \bibinfo{publisher}{Curran Associates}, \bibinfo{address}{Virtual}, \bibinfo{pages}{5199--5212}.
\newblock


\bibitem[Covington et~al\mbox{.}(2016)]%
        {covington2016deep}
\bibfield{author}{\bibinfo{person}{Paul Covington} {et~al\mbox{.}}} \bibinfo{year}{2016}\natexlab{}.
\newblock \showarticletitle{Deep neural networks for youtube recommendations}. In \bibinfo{booktitle}{\emph{RecSys}}. \bibinfo{publisher}{ACM}, \bibinfo{address}{Boston, MA, USA}, \bibinfo{pages}{191--198}.
\newblock


\bibitem[Grbovic et~al\mbox{.}(2018)]%
        {grbovic2018real}
\bibfield{author}{\bibinfo{person}{Mihajlo Grbovic} {et~al\mbox{.}}} \bibinfo{year}{2018}\natexlab{}.
\newblock \showarticletitle{Real-time personalization using embeddings for search ranking at airbnb}. In \bibinfo{booktitle}{\emph{KDD}}. \bibinfo{publisher}{ACM}, \bibinfo{address}{London, UK}, \bibinfo{pages}{311--320}.
\newblock


\bibitem[Guo et~al\mbox{.}(2017)]%
        {factorization}
\bibfield{author}{\bibinfo{person}{Huifeng Guo}, \bibinfo{person}{Ruiming Tang}, \bibinfo{person}{Yunming Ye}, \bibinfo{person}{Zhenguo Li}, {and} \bibinfo{person}{Xiuqiang He}.} \bibinfo{year}{2017}\natexlab{}.
\newblock \showarticletitle{DeepFM: A Factorization-Machine based Neural Network for CTR Prediction}. In \bibinfo{booktitle}{\emph{Proceedings of the Twenty-Sixth International Joint Conference on Artificial Intelligence}}. \bibinfo{publisher}{International Joint Conferences on Artificial Intelligence Organization}, \bibinfo{address}{Melbourne, Australia}, \bibinfo{pages}{1725--1731}.
\newblock
\href{https://doi.org/10.24963/ijcai.2017/239}{doi:\nolinkurl{10.24963/ijcai.2017/239}}


\bibitem[Huang et~al\mbox{.}(2020)]%
        {huang2020embedding}
\bibfield{author}{\bibinfo{person}{Jui-Ting Huang} {et~al\mbox{.}}} \bibinfo{year}{2020}\natexlab{}.
\newblock \showarticletitle{Embedding-based retrieval in facebook search}. In \bibinfo{booktitle}{\emph{KDD}}. \bibinfo{publisher}{ACM}, \bibinfo{address}{Virtual Event, CA, USA}, \bibinfo{pages}{2553--2561}.
\newblock


\bibitem[Huang et~al\mbox{.}(2013)]%
        {huang2013learning}
\bibfield{author}{\bibinfo{person}{Po-Sen Huang} {et~al\mbox{.}}} \bibinfo{year}{2013}\natexlab{}.
\newblock \showarticletitle{Learning deep structured semantic models for web search using clickthrough data}. In \bibinfo{booktitle}{\emph{CIKM}}. \bibinfo{publisher}{ACM}, \bibinfo{address}{San Francisco, California, USA}, \bibinfo{pages}{2333--2338}.
\newblock


\bibitem[J{\'e}gou et~al\mbox{.}(2011)]%
        {jegou2010product}
\bibfield{author}{\bibinfo{person}{Herv{\'e} J{\'e}gou}, \bibinfo{person}{Matthijs Douze}, \bibinfo{person}{Cordelia Schmid}, {and} \bibinfo{person}{Patrick P{\'e}rez}.} \bibinfo{year}{2011}\natexlab{}.
\newblock \showarticletitle{Product Quantization for Nearest Neighbor Search}.
\newblock \bibinfo{journal}{\emph{IEEE Transactions on Pattern Analysis and Machine Intelligence}} \bibinfo{volume}{33}, \bibinfo{number}{1} (\bibinfo{year}{2011}), \bibinfo{pages}{117--128}.
\newblock
\href{https://doi.org/10.1109/TPAMI.2010.57}{doi:\nolinkurl{10.1109/TPAMI.2010.57}}


\bibitem[Johnson et~al\mbox{.}(2021)]%
        {johnson2019billion}
\bibfield{author}{\bibinfo{person}{Jeff Johnson}, \bibinfo{person}{Matthijs Douze}, {and} \bibinfo{person}{Herv{\'e} J{\'e}gou}.} \bibinfo{year}{2021}\natexlab{}.
\newblock \showarticletitle{Billion-Scale Similarity Search with {GPUs}}.
\newblock \bibinfo{journal}{\emph{IEEE Transactions on Big Data}} \bibinfo{volume}{7}, \bibinfo{number}{3} (\bibinfo{year}{2021}), \bibinfo{pages}{535--547}.
\newblock
\href{https://doi.org/10.1109/TBDATA.2019.2921572}{doi:\nolinkurl{10.1109/TBDATA.2019.2921572}}


\bibitem[Khattab and Zaharia(2020)]%
        {khattab2020colbert}
\bibfield{author}{\bibinfo{person}{Omar Khattab} {and} \bibinfo{person}{Matei Zaharia}.} \bibinfo{year}{2020}\natexlab{}.
\newblock \showarticletitle{{ColBERT}: Efficient and Effective Passage Search via Contextualized Late Interaction over {BERT}}. In \bibinfo{booktitle}{\emph{SIGIR}}. \bibinfo{publisher}{ACM}, \bibinfo{address}{Virtual Event}, \bibinfo{pages}{39--48}.
\newblock
\href{https://doi.org/10.1145/3397271.3401075}{doi:\nolinkurl{10.1145/3397271.3401075}}


\bibitem[Liang et~al\mbox{.}(2023)]%
        {liang2023xlmv}
\bibfield{author}{\bibinfo{person}{Davis Liang}, \bibinfo{person}{Hila Gonen}, \bibinfo{person}{Yuning Mao}, \bibinfo{person}{Rui Hou}, \bibinfo{person}{Naman Goyal}, \bibinfo{person}{Marjan Ghazvininejad}, \bibinfo{person}{Luke Zettlemoyer}, {and} \bibinfo{person}{Madian Khabsa}.} \bibinfo{year}{2023}\natexlab{}.
\newblock \showarticletitle{{XLM}-{V}: Overcoming the Vocabulary Bottleneck in Multilingual Masked Language Models}. In \bibinfo{booktitle}{\emph{Proceedings of the 2023 Conference on Empirical Methods in Natural Language Processing}}, \bibfield{editor}{\bibinfo{person}{Houda Bouamor}, \bibinfo{person}{Juan Pino}, {and} \bibinfo{person}{Kalika Bali}} (Eds.). \bibinfo{publisher}{Association for Computational Linguistics}, \bibinfo{address}{Singapore}, \bibinfo{pages}{13142--13152}.
\newblock
\href{https://doi.org/10.18653/v1/2023.emnlp-main.813}{doi:\nolinkurl{10.18653/v1/2023.emnlp-main.813}}


\bibitem[Liu et~al\mbox{.}(2021)]%
        {liu2021que2search}
\bibfield{author}{\bibinfo{person}{Yiqun Liu} {et~al\mbox{.}}} \bibinfo{year}{2021}\natexlab{}.
\newblock \showarticletitle{Que2Search: fast and accurate query and document understanding for search at Facebook}. In \bibinfo{booktitle}{\emph{KDD}}. \bibinfo{publisher}{ACM}, \bibinfo{address}{Virtual Event}, \bibinfo{pages}{3376--3384}.
\newblock


\bibitem[Malkov and Yashunin(2020)]%
        {malkov2018efficient}
\bibfield{author}{\bibinfo{person}{Yu~A. Malkov} {and} \bibinfo{person}{Dmitry~A. Yashunin}.} \bibinfo{year}{2020}\natexlab{}.
\newblock \showarticletitle{Efficient and Robust Approximate Nearest Neighbor Search Using Hierarchical Navigable Small World Graphs}.
\newblock \bibinfo{journal}{\emph{IEEE Transactions on Pattern Analysis and Machine Intelligence}} \bibinfo{volume}{42}, \bibinfo{number}{4} (\bibinfo{year}{2020}), \bibinfo{pages}{824--836}.
\newblock
\href{https://doi.org/10.1109/TPAMI.2018.2889473}{doi:\nolinkurl{10.1109/TPAMI.2018.2889473}}


\bibitem[Peng et~al\mbox{.}(2026)]%
        {tian2026svfusion}
\bibfield{author}{\bibinfo{person}{Yuchen Peng}, \bibinfo{person}{Dingyu Yang}, \bibinfo{person}{Zhongle Xie}, \bibinfo{person}{Ji Sun}, \bibinfo{person}{Lidan Shou}, \bibinfo{person}{Ke Chen}, {and} \bibinfo{person}{Gang Chen}.} \bibinfo{year}{2026}\natexlab{}.
\newblock \showarticletitle{SVFusion: A CPU-GPU Co-Processing Architecture for Large-Scale Real-Time Vector Search}.
\newblock \bibinfo{journal}{\emph{Proceedings of the VLDB Endowment}} \bibinfo{volume}{19}, \bibinfo{number}{5} (\bibinfo{year}{2026}), \bibinfo{pages}{1074--1087}.
\newblock
\href{https://doi.org/10.14778/3796195.3796216}{doi:\nolinkurl{10.14778/3796195.3796216}}


\bibitem[Qi et~al\mbox{.}(2025)]%
        {faiss_cuvs_2025}
\bibfield{author}{\bibinfo{person}{Junjie Qi}, \bibinfo{person}{Gergely Szilvasy}, \bibinfo{person}{Michael Norris}, {and} \bibinfo{person}{Vishal Gandhi}.} \bibinfo{year}{2025}\natexlab{}.
\newblock \bibinfo{title}{Accelerating {GPU} indexes in {Faiss} with {NVIDIA} cuVS}.
\newblock \bibinfo{howpublished}{\url{https://engineering.fb.com/2025/05/08/data-infrastructure/accelerating-gpu-indexes-in-faiss-with-nvidia-cuvs/}}.
\newblock
\newblock
\shownote{Meta Engineering blog; accessed 2026-07-26}.


\bibitem[Subramanya et~al\mbox{.}(2019)]%
        {subramanya2019diskann}
\bibfield{author}{\bibinfo{person}{Suhas~Jayaram Subramanya}, \bibinfo{person}{Fnu Devvrit}, \bibinfo{person}{Harsha~Vardhan Simhadri}, \bibinfo{person}{Ravishankar Krishnaswamy}, {and} \bibinfo{person}{Rohan Kadekodi}.} \bibinfo{year}{2019}\natexlab{}.
\newblock \showarticletitle{{DiskANN}: Fast Accurate Billion-point Nearest Neighbor Search on a Single Node}. In \bibinfo{booktitle}{\emph{Advances in Neural Information Processing Systems (NeurIPS)}}. \bibinfo{publisher}{Curran Associates}, \bibinfo{address}{Vancouver, BC, Canada}, \bibinfo{pages}{13748--13758}.
\newblock
\urldef\tempurl%
\url{https://papers.nips.cc/paper/2019/hash/09853c7fb1d3f8ee67a61b6bf4a7f8e6-Abstract.html}
\showURL{%
\tempurl}


\bibitem[Sun et~al\mbox{.}(2026)]%
        {sun2025grank}
\bibfield{author}{\bibinfo{person}{Yijia Sun}, \bibinfo{person}{Shanshan Huang}, \bibinfo{person}{Zhiyuan Guan}, \bibinfo{person}{Qiang Luo}, \bibinfo{person}{Ruiming Tang}, \bibinfo{person}{Kun Gai}, {and} \bibinfo{person}{Guorui Zhou}.} \bibinfo{year}{2026}\natexlab{}.
\newblock \showarticletitle{GRank: Towards Target-Aware and Streamlined Industrial Retrieval with a Generate-Rank Framework}. In \bibinfo{booktitle}{\emph{Proceedings of the ACM Web Conference 2026}}. \bibinfo{publisher}{ACM}, \bibinfo{address}{New York, NY, USA}, \bibinfo{pages}{7798--7808}.
\newblock
\href{https://doi.org/10.1145/3774904.3792810}{doi:\nolinkurl{10.1145/3774904.3792810}}


\bibitem[Tay et~al\mbox{.}(2022)]%
        {tay2022transformer}
\bibfield{author}{\bibinfo{person}{Yi Tay}, \bibinfo{person}{Vinh~Q. Tran}, \bibinfo{person}{Mostafa Dehghani}, \bibinfo{person}{Jianmo Ni}, \bibinfo{person}{Dara Bahri}, \bibinfo{person}{Harsh Mehta}, \bibinfo{person}{Zhen Qin}, \bibinfo{person}{Kai Hui}, \bibinfo{person}{Zhe Zhao}, \bibinfo{person}{Jai Gupta}, \bibinfo{person}{Tal Schuster}, \bibinfo{person}{William~W. Cohen}, {and} \bibinfo{person}{Donald Metzler}.} \bibinfo{year}{2022}\natexlab{}.
\newblock \showarticletitle{Transformer Memory as a Differentiable Search Index}. In \bibinfo{booktitle}{\emph{Advances in Neural Information Processing Systems}}, Vol.~\bibinfo{volume}{35}. \bibinfo{publisher}{Curran Associates, Inc.}, \bibinfo{address}{Red Hook, NY, USA}, \bibinfo{pages}{21831--21843}.
\newblock


\bibitem[Tian et~al\mbox{.}(2024)]%
        {tian2024fusionanns}
\bibfield{author}{\bibinfo{person}{Bing Tian}, \bibinfo{person}{Haikun Liu}, \bibinfo{person}{Yuhang Tang}, \bibinfo{person}{Shihai Xiao}, \bibinfo{person}{Zhuohui Duan}, \bibinfo{person}{Xiaofei Liao}, \bibinfo{person}{Xuecang Zhang}, \bibinfo{person}{Junhua Zhu}, {and} \bibinfo{person}{Yu Zhang}.} \bibinfo{year}{2024}\natexlab{}.
\newblock \showarticletitle{FusionANNS: An Efficient CPU/GPU Cooperative Processing Architecture for Billion-scale Approximate Nearest Neighbor Search}.
\newblock \bibinfo{journal}{\emph{arXiv preprint arXiv:2409.16576}}  \bibinfo{volume}{abs/2409.16576} (\bibinfo{year}{2024}), \bibinfo{pages}{1--15}.
\newblock


\bibitem[Wang et~al\mbox{.}(2018)]%
        {wang2018billion}
\bibfield{author}{\bibinfo{person}{Jizhe Wang}, \bibinfo{person}{Pipei Huang}, \bibinfo{person}{Huan Zhao}, \bibinfo{person}{Zhibo Zhang}, \bibinfo{person}{Binqiang Zhao}, {and} \bibinfo{person}{Dik~Lun Lee}.} \bibinfo{year}{2018}\natexlab{}.
\newblock \showarticletitle{Billion-scale Commodity Embedding for E-commerce Recommendation in Alibaba}. In \bibinfo{booktitle}{\emph{Proceedings of the 24th ACM SIGKDD International Conference on Knowledge Discovery \& Data Mining}}. \bibinfo{publisher}{ACM}, \bibinfo{address}{London, UK}, \bibinfo{pages}{839--848}.
\newblock
\href{https://doi.org/10.1145/3219819.3219869}{doi:\nolinkurl{10.1145/3219819.3219869}}


\bibitem[Wang et~al\mbox{.}(2024)]%
        {wang2024improving}
\bibfield{author}{\bibinfo{person}{Liang Wang}, \bibinfo{person}{Nan Yang}, \bibinfo{person}{Xiaolong Huang}, \bibinfo{person}{Linjun Yang}, \bibinfo{person}{Rangan Majumder}, {and} \bibinfo{person}{Furu Wei}.} \bibinfo{year}{2024}\natexlab{}.
\newblock \showarticletitle{Improving Text Embeddings with Large Language Models}. In \bibinfo{booktitle}{\emph{Proceedings of the 62nd Annual Meeting of the Association for Computational Linguistics (ACL)}}. \bibinfo{publisher}{Association for Computational Linguistics}, \bibinfo{address}{Bangkok, Thailand}, \bibinfo{pages}{11897--11916}.
\newblock
\href{https://doi.org/10.18653/v1/2024.acl-long.642}{doi:\nolinkurl{10.18653/v1/2024.acl-long.642}}


\bibitem[Xu et~al\mbox{.}(2023)]%
        {xu2023spfresh}
\bibfield{author}{\bibinfo{person}{Yuming Xu}, \bibinfo{person}{Hengyu Liang}, \bibinfo{person}{Jin Li}, \bibinfo{person}{Shuotao Xu}, \bibinfo{person}{Qi Chen}, \bibinfo{person}{Qianxi Zhang}, \bibinfo{person}{Cheng Li}, \bibinfo{person}{Ziyue Yang}, \bibinfo{person}{Fan Yang}, \bibinfo{person}{Yuqing Yang}, \bibinfo{person}{Peng Cheng}, {and} \bibinfo{person}{Mao Yang}.} \bibinfo{year}{2023}\natexlab{}.
\newblock \showarticletitle{SPFresh: Incremental in-place update for billion-scale vector search}. In \bibinfo{booktitle}{\emph{Proceedings of the 29th Symposium on Operating Systems Principles (SOSP)}}. \bibinfo{publisher}{ACM}, \bibinfo{address}{Koblenz, Germany}, \bibinfo{pages}{545--561}.
\newblock
\href{https://doi.org/10.1145/3600006.3613166}{doi:\nolinkurl{10.1145/3600006.3613166}}


\bibitem[Xue et~al\mbox{.}(2026)]%
        {silvertorch}
\bibfield{author}{\bibinfo{person}{Bi Xue} {et~al\mbox{.}}} \bibinfo{year}{2026}\natexlab{}.
\newblock \showarticletitle{SilverTorch: A Unified Model-based System to Democratize Large-Scale Recommendation on GPUs}. In \bibinfo{booktitle}{\emph{Proceedings of the 49th International ACM SIGIR Conference on Research and Development in Information Retrieval}}. \bibinfo{publisher}{ACM}, \bibinfo{address}{New York, NY, USA}, \bibinfo{pages}{2139--2149}.
\newblock
\href{https://doi.org/10.1145/3805712.3809755}{doi:\nolinkurl{10.1145/3805712.3809755}}


\bibitem[Ying et~al\mbox{.}(2018)]%
        {ying2018graph}
\bibfield{author}{\bibinfo{person}{Rex Ying} {et~al\mbox{.}}} \bibinfo{year}{2018}\natexlab{}.
\newblock \showarticletitle{Graph convolutional neural networks for web-scale recommender systems}. In \bibinfo{booktitle}{\emph{KDD}}. \bibinfo{publisher}{ACM}, \bibinfo{address}{London, UK}, \bibinfo{pages}{974--983}.
\newblock


\bibitem[Zhu et~al\mbox{.}(2025)]%
        {li2023llm}
\bibfield{author}{\bibinfo{person}{Yutao Zhu}, \bibinfo{person}{Huaying Yuan}, \bibinfo{person}{Shuting Wang}, \bibinfo{person}{Jiongnan Liu}, \bibinfo{person}{Wenhan Liu}, \bibinfo{person}{Chenlong Deng}, \bibinfo{person}{Haonan Chen}, \bibinfo{person}{Zheng Liu}, \bibinfo{person}{Zhicheng Dou}, {and} \bibinfo{person}{Ji-Rong Wen}.} \bibinfo{year}{2025}\natexlab{}.
\newblock \showarticletitle{Large Language Models for Information Retrieval: A Survey}.
\newblock \bibinfo{journal}{\emph{ACM Transactions on Information Systems}} \bibinfo{volume}{44}, \bibinfo{number}{1} (\bibinfo{year}{2025}), \bibinfo{pages}{1--54}.
\newblock
\href{https://doi.org/10.1145/3748304}{doi:\nolinkurl{10.1145/3748304}}


\end{thebibliography}

\appendix
\section{Supplementary Definitions and Measurement Details}

\subsection{Per-Pathway Serving Footprint}
\label{app:hardware}

\begin{table}[H]
\centering
\small
\setlength{\tabcolsep}{2pt}
\caption{Serving scopes. The GPU benchmark is per accelerator and excludes end-to-end search stages.}
\label{tab:hardware}
\begin{tabular}{@{}>{\raggedright\arraybackslash}p{0.20\columnwidth} >{\raggedright\arraybackslash}p{0.30\columnwidth} >{\raggedright\arraybackslash}p{0.36\columnwidth}@{}}
\toprule
 & GPU pathway & CPU pathway \\
\midrule
Hardware & MI300X accelerator & Commodity x86 DRAM hosts \\
Inventory & Billion-scale selected pool & Tens-of-billions-scale inventory \\
Refresh & Daily model + snapshot & Base backfill + live updates \\
Measurement & Roughly 150--200 queries/s/card; 99th-percentile latency $\sim$30--40 ms & Online CPU-use measurements \\
Excludes & Network, aggregation, downstream ranker & No comparable per-host latency reported \\
\bottomrule
\end{tabular}
\end{table}

The scopes in Table~\ref{tab:hardware} must not be conflated. The per-card tens-of-milliseconds value characterizes the GPU model server. Table~\ref{tab:retrieval_latency} measures the wider retrieval-branch execution block for each pathway over posts, including its in-process orchestration. Neither measurement includes the complete search request through downstream ranking and client rendering.

\subsection{Overlap and Shared-Ranker Input Composition}
\label{app:overlap}

For each search request $i$ for which both dense pathways ran, let $G_i$ be the GPU set and $C_i$ the union of the three dense CPU retrieval sets. We compute
\begin{align}
O_G(i)&=\frac{|G_i\cap C_i|}{|G_i|},\qquad
O_C(i)=\frac{|G_i\cap C_i|}{|C_i|},\\
J(i)&=\frac{|G_i\cap C_i|}{|G_i\cup C_i|},
\end{align}
then average each ratio over requests in a popularity segment. Requests in which only one pathway runs are excluded because the question is redundancy when both run. The main result uses a three-day window. A separate seven-day computation gives GPU-side overlap of $2.32\%$, $2.07\%$, and $2.15\%$ for head, torso, and tail, within 0.13 percentage points of the main window.

\begin{table}[H]
\centering
\small
\setlength{\tabcolsep}{4pt}
\caption{Deduplicated composition at the shared-ranker input per eligible search request. Counts are independently rounded; ``Overlap'' contains documents attributed to more than one group.}
\label{tab:final_pool}
\begin{tabular}{@{}lrr@{}}
\toprule
Exclusive source group & Candidates & Share \\
\midrule
GPU only & 209 & $27\%$ \\
Dense CPU only & 47 & $6\%$ \\
Lexical CPU only & 500 & $64\%$ \\
Cross-group overlap & 26 & $3\%$ \\
\midrule
Total & $\sim$781 & $100\%$ \\
\bottomrule
\end{tabular}
\end{table}

These are candidate counts at the shared-ranker input, not final Top-20 impressions. They therefore support source diversity but not per-source exposure or engagement claims.

\subsection{Training Objectives by Pathway}
\label{app:objectives}

\begin{table}[H]
\centering
\small
\setlength{\tabcolsep}{3pt}
\caption{Training objectives and serving roles.}
\label{tab:training_objectives}
\begin{tabular}{@{}p{0.11\columnwidth}p{0.25\columnwidth}p{0.54\columnwidth}@{}}
\toprule
Path & Objective & Supervision and serving role \\
\midrule
GPU & Cross-session InfoNCE & In-batch documents from other sessions; learns the ANN space. \\
GPU & Within-session InfoNCE & Engaged positives and relevant non-engaged negatives. \\
GPU & Binary cross-entropy + Smooth L1 & Engagement and relevance targets for interaction scoring. \\
GPU & Weighted value model & Task weights select an operating point. \\
CPU & Bidirectional InfoNCE & Produces reusable 384-dimensional document vectors. \\
\bottomrule
\end{tabular}
\end{table}

\subsection{Capacity-Plan Assumptions}
\label{app:tco}

Table~\ref{tab:tco_main} concerns a $d=256$ SQ8 representation, not the deployed CPU model's 384-dimensional representation. For this broad-vector workload, the accelerator provides roughly three to four times the per-unit query throughput and two to three times the per-unit vector capacity of a commodity CPU host. The resulting storage-dominated plan uses roughly one-third as many accelerator units per region. These ratios are distinct from the interaction-model benchmark in Section~\ref{sec:gpu_serving}. The CPU plan includes modest inventory headroom, making the reported accelerator/CPU cost ratio conservative relative to storage-minimal CPU provisioning. The source does not report utilization, a paired latency target, or a recall/quality-equivalence test, and the plan does not include the complete hybrid stack.

\subsection{Measurement Procedure and Reproducibility}
\label{app:repro}

The primary result is the relative change from a production A/B against the legacy all-CPU control. The experiment uses persistent account-level assignment, split approximately equally between control and treatment, and includes millions of assigned accounts per arm per day. Both headline metrics are aggregated identically over the same nine consecutive days. Because the same accounts can recur across days, we do not divide a daily standard error by $\sqrt{9}$. The Cauchy--Schwarz bound permits arbitrary cross-day covariance and upper-bounds the aggregate 95\% half-width by the mean of the nine daily half-widths, taking the larger reported side for each day. For GSRR, this procedure yields the conservative 95\% confidence interval $[+1.71\%,+2.32\%]$ around an equal-day mean lift of $+2.01\%$; for DCG@20 it yields $[+3.99\%,+5.04\%]$ around $+4.51\%$. Every daily interval of both metrics is above zero. The GPU-pathway experiment and GPU-disabled CPU-pathway evaluation ran in separate windows against their contemporaneous production controls and report their own two-sided 95\% intervals; they are not pooled with the full-system test. Overlap is computed per eligible request from a later log window and averaged by popularity.

The manuscript provides the equations, treatment scopes, aggregation procedures, and aggregate tables. Production configurations and privacy-sensitive logs remain unavailable.

\end{document}